\documentclass{aa}  

\newcommand\mJybeam{mJy beam$^{-1}$}

\newcommand{\Mm}		{$\pm$}

\usepackage{graphicx}
\usepackage{xcolor}
\usepackage{ulem}
\usepackage{cancel}
\usepackage{txfonts}
\begin{document}

   \title{Magnetic field in a circumbinary disk around a Class I YSO}

\author{F.~O. Alves
\inst{1}\fnmsep\thanks{Corresponding author: Felipe O. Alves (falves@mpe.mpg.de)}
\and
J.~M. Girart \inst{2,3}
\and
M. Padovani\inst{4}
\and
D. Galli\inst{4}
\and
G.~A.~P.  Franco\inst{5}
\and
P. Caselli\inst{1}
\and
W.~H.~T. Vlemmings\inst{6}
\and
Q. Zhang\inst{7}
\and
H. Wiesemeyer\inst{8}
} 

\institute{Max-Planck-Institut f\"ur extraterrestrische Physik, 
Giessenbachstr. 1, 
D-85748 Garching, Germany
\and
Institut de Ci\`encies de l'Espai (ICE, CSIC), 
Can Magrans S/N,
E-08193 Cerdanyola del Vall\`es, Catalonia, Spain
\and
Institut d'Estudis Espacials de de Catalunya (IEEC),
 E-08034 Barcelona, Catalonia, Spain
\and
INAF-Osservatorio Astrofisico di Arcetri, 
Largo E. Fermi 5,
50125 Firenze, Italy
\and
Departamento de F\'isica--ICEx--UFMG, 
Caixa Postal 702,
30.123-970 Belo Horizonte, Brazil
\and
Department of Earth and Space Sciences, Chalmers University of Technology,
Onsala Space Observatory,
439 92 Onsala, Sweden
\and
Harvard-Smithsonian Center for Astrophysics, 
60 Garden Street,
MA 02138 Cambridge, USA
\and
Max-Planck-Institut f\"ur Radioastronomie, 
Auf dem H\"ugel 69
53121 Bonn, Germany
}

   \date{Received ; accepted \today}

 
  \abstract
   {Polarized continuum emission at millimeter/sub-millimeter wavelengths is usually attributed to thermal emission from 
   dust grains aligned through radiative torques with the magnetic field. However, recent theoretical work has shown that 
   under specific conditions polarization may arise from self-scattering of thermal emission and by radiation fields from a 
   nearby stellar object.}
   {We use multi-frequency polarization observations of a circumbinary disk to investigate how the polarization properties
   change at distinct frequency bands. Our goal is to discern the main mechanism responsible for the polarization through 
   comparison between our observations and model predictions for each of the proposed mechanisms.}   
   {We used the Atacama Large Millimeter/submillimeter Array to perform full polarization observations at 97.5 GHz (Band 
   3), 233 GHz (Band 6) and 343.5 GHz (Band 7). The ALMA data have a mean spatial resolution of 28 AU. The target is 
   the Class I object BHB07-11, which is the youngest object in the Barnard 59 protocluster. Complementary Karl G. 
   Jansky Very Large Array observations at 34.5 GHz were also performed and revealed a binary system at centimetric 
   continuum emission within the disk.}
   {We detect an extended and structured polarization pattern remarkably consistent among all three bands. The 
   distribution of polarized intensity resembles a horseshoe shape with polarization angles following this morphology. 
   From the spectral index between bands 3 and 7, we derive a dust opacity index $\beta \sim 1$ 
   consistent with maximum grain sizes larger than expected to produce self-scattering polarization in each band. The 
   polarization morphology and the polarization levels do not match predictions from self-scattering. On 
   the other hand, marginal correspondence is seen between our maps and predictions from radiation field assuming the  
   brightest binary component as main radiation source. Previous molecular line data from BHB07-11 indicates disk 
   rotation. We used the DustPol module of the ARTIST radiative transfer tool to produce synthetic polarization maps from 
   a rotating magnetized disk model assuming combined poloidal and toroidal magnetic field components. The magnetic
   field vectors (i. e., the polarization vectors rotated by 90$\degr$) are better represented by a model with poloidal magnetic
   field strength about 3 times the toroidal one.}
   {The similarity of our polarization patterns among the three bands provides a strong evidence against 
   self-scattering and radiation fields. On the other hand, our data are reasonably well reproduced by a model of disk with 
   toroidal magnetic field components slightly smaller than poloidal ones. The residual is likely due to the 
   internal twisting of the magnetic field due to the binary system dynamics, which is not considered in our model.}


   \maketitle
%

\section{Introduction}

Polarized emission of  dust grains is considered a good tracer of the magnetic field in the interstellar medium, since 
aspherical dust particles tend to align their short axis parallel to the local direction of the field, with the help of radiative 
torques \citep[see, e.\,g., ][]{Draine97, Andersson15}. The method has been successfully used to map the morphology of 
the magnetic field in star forming regions, over scales ranging from molecular clouds \citep{PlanckXXXV16} to protostellar 
envelopes \citep[e.\,g.,][]{Girart06,Alves11a,Hull14,Zhang14}. Recently, millimeter-wave polarimetry has become sensitive 
enough to allow to detect the polarized emission of dust in protostellar and protoplanetary disks 
\citep[starting from][]{Rao14}, opening the way to address many open questions concerning the role of magnetic fields in 
the formation and evolution of circumstellar disks \citep{Lizano15}. With the Atacama Large Millimeter/submillimeter Array
(ALMA), our knowledge on the polarization properties of young stellar objects (YSO) has increased substantially due
to the high sensitivity and angular resolution achievable by this instrument \citep[e.g., ][]{Hull17a,Hull17b,Lee18,Cox18}.

However, the interpretation of millimeter-wave polarization observations of disks is not straightforward, because other 
processes than just emission by magnetically aligned grains may be at play in environments where a significant amount 
of large grains is expected to be produced by dust coagulation and the local radiative field is relatively intense. In fact, 
polarized emission in disks can also be produced by self-scattering of an anisotropic radiation field 
\citep{Kataoka15,Yang16a,Yang16b} and alignment with the radiation anisotropy \citep[also known as ``radiative 
alignment'', ][]{Lazarian07b,Tazaki17}. The former prevails over 
magnetic alignment if sufficiently large grains are present, resulting in a high scattering opacity and therefore a high 
contribution of scattered light in the observed emission; the latter dominates if the grain precession rate induced by 
radiative torques is faster than the Larmor precession around the magnetic field.

In the case of self-scattering, the polarization angle and degree depend on the grain properties and the system geometry: 
for Rayleigh scattering, the polarization pattern is centro-symmetric for face-on disks, and parallel to the minor axis for 
inclined disks; in the case of radiative alignment, the polarization is perpendicular to the propagation direction of the 
radiative flux. In this context, the polarized emission observed at three millimeter-wave bands in the HL Tau disk has been 
interpreted as due to self-scattering at 870~$\mu$m, and radiative alignment at 3~mm, with intermediate characteristics 
at 1.3~mm \citep{Kataoka17,Stephens17}. This example stresses the importance of comparing maps at different 
wavelengths in order to disentangle the possible different contributions to the observed polarization.

This paper introduces the results obtained from the polarized emission observed at three millimeter/sub-millimeter-wave bands in the BHB07-11 disk, a Class I object embedded in the darkest parts of the Pipe Nebula, a quite quiescent complex of 
interstellar clouds located at a distance of 145\,pc from the Sun \citep{Alves07}. Only a handful of embedded sources are
known in the entire complex, among which BHB07-11 is the youngest member of a protocluster of low-mass YSOs
\citep[e.g.,][]{Brooke07}.

\begin{figure*}[t!]
\includegraphics[width=\textwidth, trim={1cm 0 1.8cm 0}]{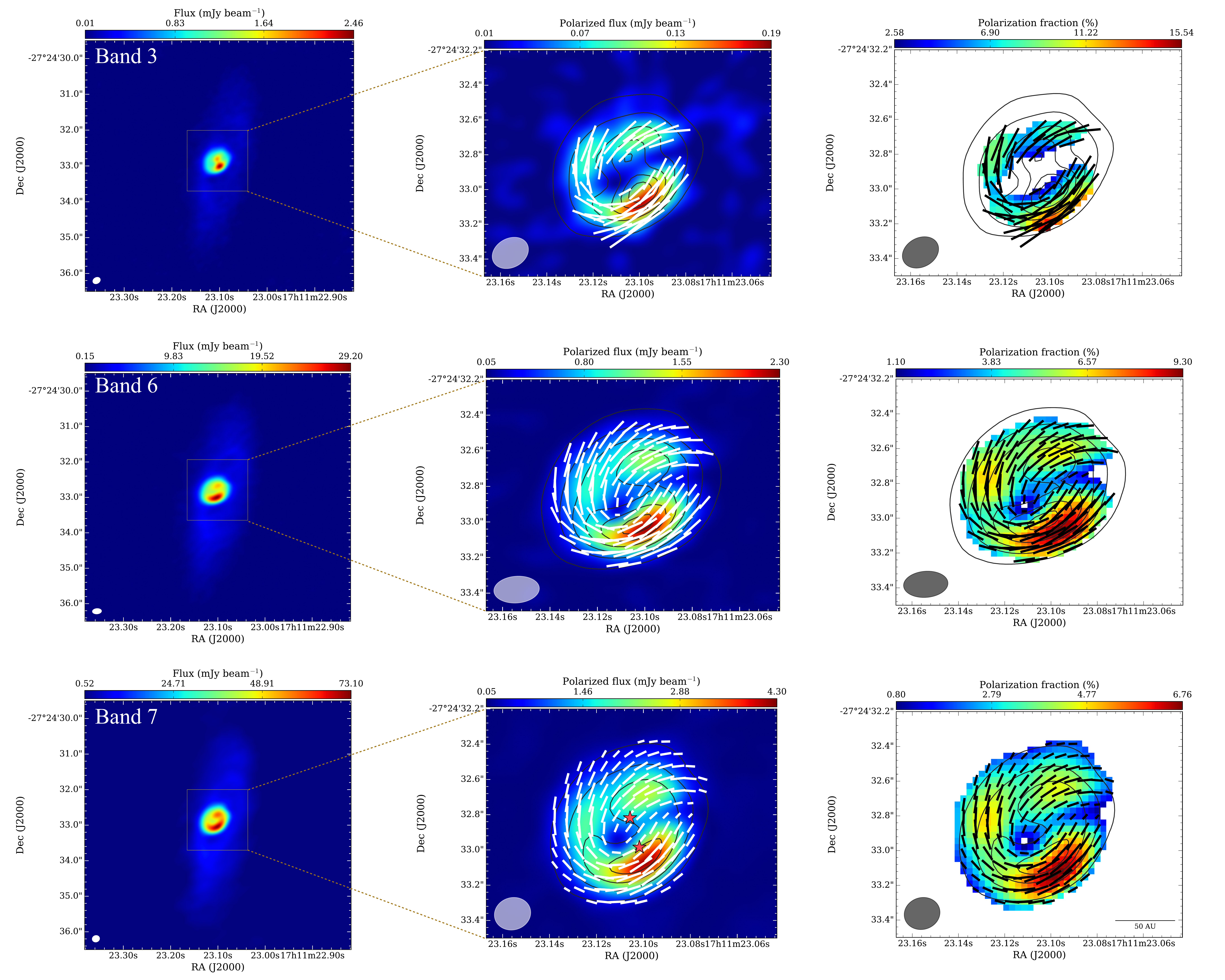}
\caption{Stokes I (left panel), polarized intensity (middle panel) and polarization fraction (right panel) of our Band 3 
(upper panels), Band 6 (middle panels) and Band 7 (lower panels) ALMA data. The contours indicate the intensity levels 
of the Stokes I emission in each band, being 30, 60, 90, 120, 150,180 times the {\it rms} for bands of 3 and 6 
({\it rms}$_{\mathrm{Band~3}} \sim 0.012$ mJy beam$^{-1}$ and {\it rms}$_{\mathrm{Band~6}} \sim 0.15$ mJy beam
$^{-1}$), and 30, 60, 90, 120 times 0.52 mJy beam$^{-1}$, the {\it rms} for Band 7. The synthesized beams are $0.22\arcsec \times 0.17\arcsec$ (position angle = $-61.46\degr$) for Band 3, $0.26\arcsec \times 0.15\arcsec$ (position angle = $-85.65\degr$) for Band 6 and $0.21\arcsec \times 0.18\arcsec$ (position angle = $-72.79\degr$) for Band 7. The vector map indicates the 
orientation of the polarization whose length is proportional to the polarization fraction. The vectors are sampled as one 
vector every two pixels. The two star-symbols at the middle-lower panel indicate the position of the two continuum peaks 
detected in the 34 GHz VLA data, being VLA 5a the northern peak and VLA 5b the southern one (see also Fig. \ref{fig:vla+alma}). 
\label{fig:cuts}}
\end{figure*}

In an earlier non-polarimetric study of BHB07-11 it has been reported the detection of a bipolar molecular outflow 
launched beyond the edge of the disk, at a distance of $\sim$90--130 astronomical units (AU) from the rotational axis, 
where supposedly infalling gas lands on the disk \citep{Alves17}. In this paper, we revisit this object in order to investigate 
the characteristics of the millimeter polarized emission from its disk.
 
\section{Observations}
\label{sec:obs}

\subsection{ALMA}
\label{subsec:alma}

We used the ALMA in full polarization mode, where all correlations 
between the linear feeds (X, Y) of each antenna are processed. These correlations include the parallel-hand (XX, YY) 
and the cross-hand (XY, YX) visibilities from which we obtain the Stokes parameters I, Q and U in the image space.  
The spectral setup have four spectral windows in Time Domain Mode (TDM, 64 channels with a width of 31.5 kHz), 
which is optimized for observations of continuum emission. We used the atmospheric windows centered at 97.5 GHz 
(Band 3), 233 GHz (Band 6) and 343.5 GHz (Band 7). 

The Band 3 observations were performed in two Execution Blocks (EBs) on November 14, 2017. The first EB was $\sim 
1.7$ hour long and used 44 antennas from the main array. The second EB was $\sim 2.1$ hours long and used 46 
antennas from the main array. The polarization (and flux) calibrator J1733-1304 was measured to have a peak of 
polarized intensity of 23.13 $\pm$ 0.02 \mJybeam and mean position angle of -48.8\degr (East of North, as for all references to polarization position angle from now on). Although the noise level in the image plane of the calibrator is very low, the uncertainty on the absolute flux density scale is expected to be $\sim 10$\%. The mean polarization fraction 
of $2.4$\% is consistent with earlier measurements of this object according to the ALMA Calibrator Source Catalogue. 
The phase calibrator J1700-2610 was observed to have a flux density of $1.6 \pm 0.2$ Jy, which is consistent with the nearest monitoring observations of this object according to the ALMA catalogue. Bandpass calibration was performed through observations of
J1924-2914.

The Band 6 observations  were performed on the 19$^{\mathrm{th}}$ September of 2015 using 35 antennas of the 
array. The total observing time was 3 hours divided in three EBs of 1 hour each. The absolute flux calibration was 
obtained with observing scans on Titan, the bandpass calibration with J1924-2914, phase calibration with J1713-2658 
and the polarization calibration with J1751+0939. While bandpass and gain calibration were performed individually in 
each EB, the absolute flux calibration obtained for the first EB was bootstrapped to the phase calibrator in the second and 
third EBs. After concatenation of the three EBs, the polarization calibration was performed and the leakage terms were 
computed. The peak of polarized intensity of the calibrator is $\sim 55$ mJy beam$^{-1}$, with a constant polarization 
level of $\sim 5.5$\% and position angle of $\sim 5.3\degr$. Despite 
the source variability at this band (according to the ALMA Calibrator Catalogue), the observed polarization levels are fairly 
consistent with the values observed during its monitoring. The phase calibrator J1713-2658 was measured to have a flux 
density of $0.14 \pm 0.01$ Jy. No entries are found in the ALMA Calibrator Catalogue for this band that would allow for a 
flux comparison with respect to the ALMA monitoring observations.

The Band 7 observations were carried out on the 17$^{\mathrm{th}}$ May of 2017 using 45 antennas and two EBs of 2 
hours and 1.5 hours, respectively.  Each EB was calibrated individually in (amplitude and phase) gains and bandpass, 
while the concatenated dataset was used for polarization calibration due to the optimized parallactic angle coverage. For 
these observations, the flux calibrator, J1733-1304, was also used as polarization calibrator. The flux density obtained for 
the phase calibrator, $0.55 \pm 0.06$ Jy, is consistent with the ALMA Calibrator Catalogue. The polarization calibrator 
has a peak of polarized flux of 15.9 \mJybeam~and an {\it rms} noise level of 0.05 \mJybeam. The polarized level is $\sim 1.6\%$ and position angle $59\degr$, consistent with previous observations listed in the 
ALMA catalogue.

This work relies on a multi-frequency analysis of polarization properties. Therefore, in order to ensure that the same spatial scales are being used for comparison in the distinct frequency data, we produced cleaned maps using a 
{\it uv} range that is common for all bands, from 27 to 1310 k$\lambda$. The inverse Fourier transform of the {\it uv} 
visibilities produced Stokes parameters (I, Q and U) maps with synthesized beams of $0.22\arcsec \times 0.17\arcsec$ 
(position angle = $-61.46\degr$), $0.26\arcsec \times 0.15\arcsec$ (position angle = $-85.65\degr$) and $0.21\arcsec 
\times 0.18\arcsec$ (position angle = $-72.79\degr$) for bands 3, 6 and 7, respectively. This corresponds to a mean 
spatial resolution of $\sim28.3$ AU. The total intensity (Stokes I) maps have a {\it rms} noise level of 0.012, 0.15 and 
0.52 \mJybeam~for bands 3, 6 and 7, respectively. The Stokes Q and U maps reached a noise level of $\sim 0.008$ 
\mJybeam~in Band 3 and $\sim 0.03$ \mJybeam~in Bands 6 and 7. The polarized intensity, computed as $\sqrt{Q^2 + 
U^2}$, reached a noise level of $\sim 0.01$ \mJybeam~in Band 3 and $\sim 0.05$ \mJybeam~in Bands 6 and 7. For the 
present work, we adopt a signal-to-noise ratio of 5 in polarized intensity for all bands in order to produce vector maps.

We applied primary beam correction to account for the decreasing instrument sensitivity away from the center of the 
primary beam. Our source is well centered in the phase center and no extended polarized emission is 
detected at scales larger than $5.7^{\prime\prime}$, which corresponds to $\sim \frac{1}{3} \times$ of the primary beam 
in Band 7 (the smaller of the three bands). 

\subsection{VLA}
\label{subsec:vla}

We used the Karl G. Jansky Very Large Array (VLA) to observe the continuum emission at 34.5 GHz ($\lambda \sim 
8.7$ mm, Ka band) in its most extended configuration. The observations were carried out on December 4 and 10, 2016. 
The total on-source time was 77 minutes.  The calibration was performed using the standard CASA VLA pipeline.  The 
maps were obtained from the calibrated visibilities using a robust weighting of 0, which provided a 
synthesized beam of $0.10^{\prime\prime} \times0.04^{\prime\prime}$ with a position angle of 12\degr (i.e. a spatial 
resolution of $\sim$10 AU at the Pipe nebula distance). The {\it rms} noise of the map is 12 $\mu$Jy~beam$^{-1}$.

Previous VLA lower angular resolution observations ($\sim 3^{\prime\prime}$) at 3.6 cm detected the source VLA 5 near
the center of the disk of BHB07-11 \citep{Dzib13}. The new observations resolved VLA 5 in a binary system, VLA 5a (RA = 
17:11:23.1057, DEC = $-27$:24:32.818) and VLA 5b, (17:11:23.1017, $-27$:24:32.985). The uncertainties in the positions 
are $\simeq$4~mas. Their flux densities are $0.32\pm0.03$ and $0.23\pm0.03$ mJy, respectively. The two radio sources 
appear inside the disk, slightly offset of the dust millimeter continuum peaks (Fig. \ref{fig:vla+alma}). 

\section{Polarization properties}
\label{sec:res}

The Stokes I continuum emission in the three ALMA bands is consistent with the 1.3 mm continuum map obtained in non-
polarimetric mode by  \citet{Alves17}. The thermal emission shows the inner envelope, i. e., the  elongated and tenuous 
structure that surrounds an at least 10 times brighter structure associated with a compact disk of radius $\sim 80$ AU. 
The disk brightness shows a strong asymmetry, with enhanced emission toward the southwest. 

\begin{figure}[t]
\includegraphics[width=\columnwidth]{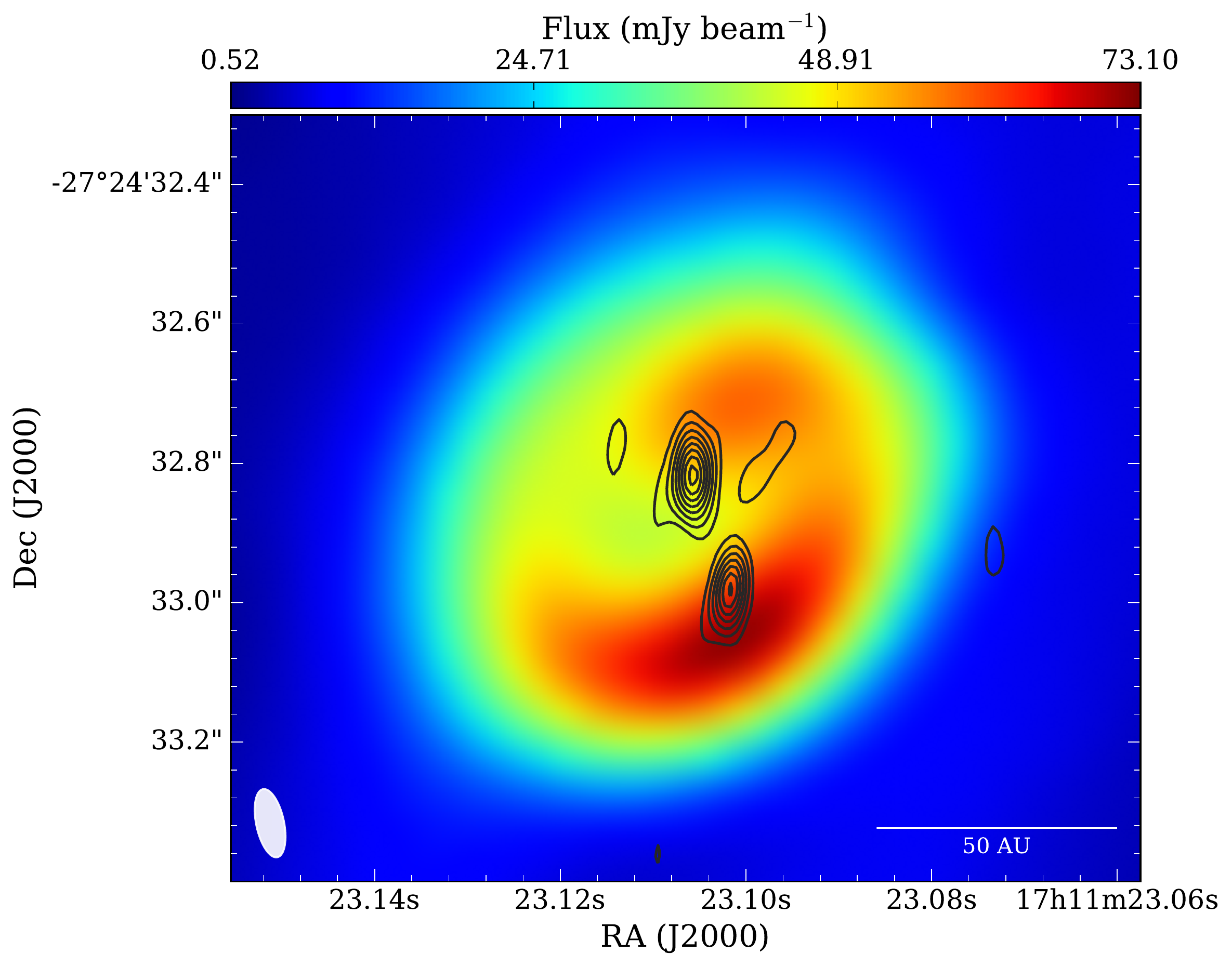}
\caption{Intensity contours of the 34 GHz VLA emission plotted on top of the Band 7 Stokes I  image of BHB07-11. The 
contours correspond to intensity levels of 3, 5, 7, ... 15, 17 $\times$ the noise level of the VLA map, 12 $\mu$Jy beam$^{-1}$.
The VLA synthesized beam of 0.10$\arcsec \times 0.04\arcsec$ is shown in the lower left corner.
\label{fig:vla+alma}}
\end{figure}

\begin{figure}[t!]
\includegraphics[width=\columnwidth]{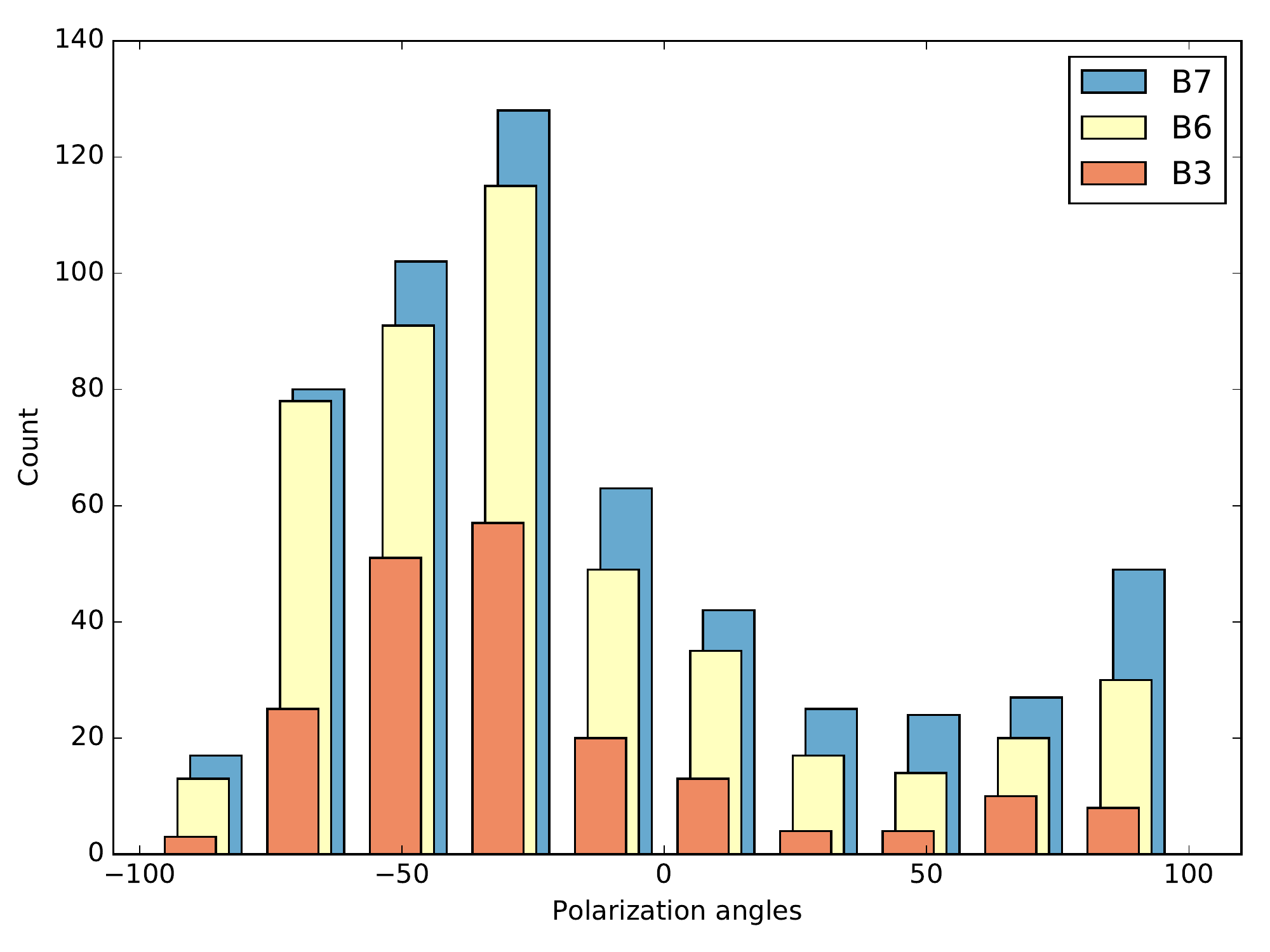}
\caption{Distribution of polarization angles for each band using a bin of 10\degr.
\label{fig:hist}}
\end{figure}

\begin{figure}[t!]
\includegraphics[width=\columnwidth]{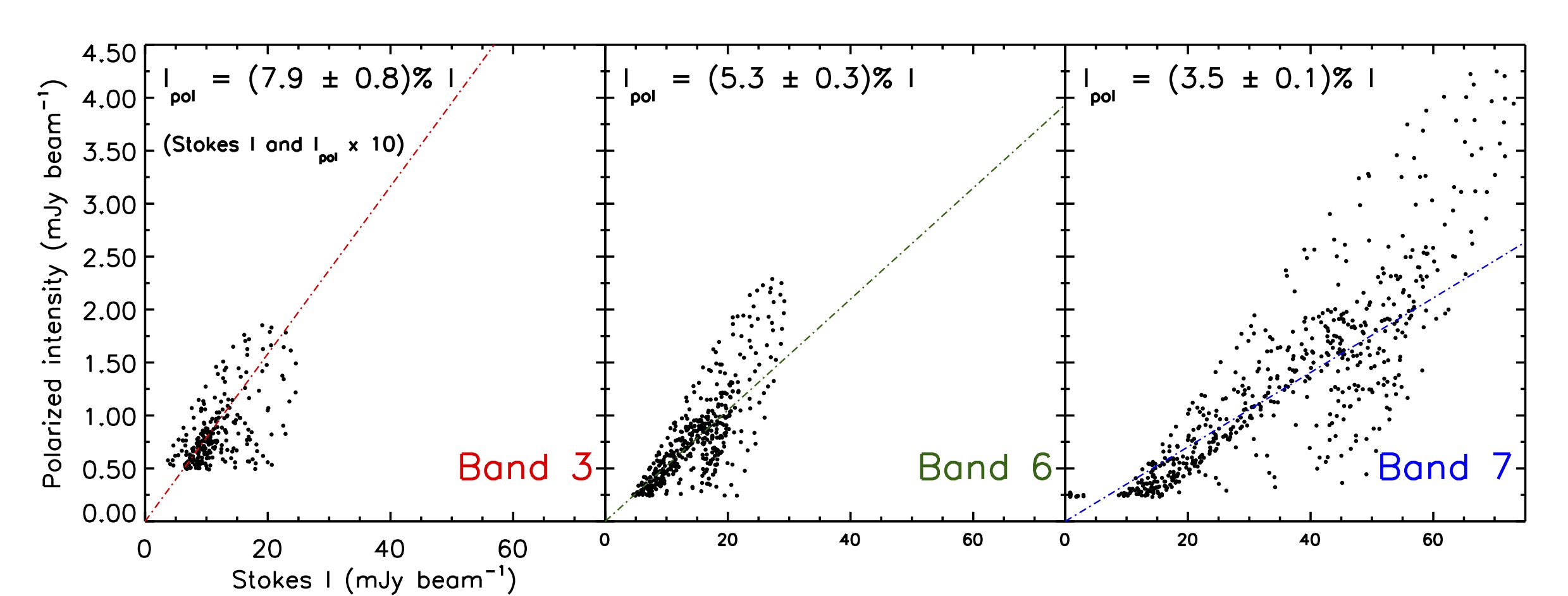}
\caption{Distribution of the polarized intensity as a function of total intensity. The mean polarization fraction over the 
regions shown in the right panels of Fig. \ref{fig:cuts} are indicated in each panel as straight lines. The {\it rms} (error 
bars) for the polarized intensity and total intensity are mentioned in \S~\ref{subsec:alma}.  
\label{fig:poli}}
\end{figure}

The polarization properties of the three bands are remarkably similar (Fig. \ref{fig:cuts}). The polarized intensity is 
spatially confined to the Stokes I emission of the disk. The distribution of polarized intensity and polarization fraction  
resembles a horseshoe shape peaking in the southwest, similar to the Stokes I emission. The 34 GHz VLA continuum 
contours are shown with respect to the Stokes I Band 7 emission in Fig. \ref{fig:vla+alma}. The polarization position angles (PA) 
have a very similar distribution in all bands (Fig. \ref{fig:hist}), with the electric field orientation following the structure of  
the polarized intensity. Interestingly, the polarized intensity increases with total intensity (Fig. \ref{fig:poli}). 
Although the  polarized intensity is systematically larger for Band 7, the polarization fraction 
($I_{\mathrm{pol}}/I_{\mathrm{total}}$) observed for Band 3 is on average larger than the polarization levels at Band 6 
and 7. Thus, the mean polarization fraction is 7.9\Mm0.8, 5.3\Mm0.3 and 3.5\Mm0.1\% for bands 3, 6 and 7, respectively. 
Yet, the spatial distribution of the polarization fraction is very similar in all three bands (right panels of Fig. \ref{fig:cuts}). 
The polarization fraction exhibit three peaks, with the strongest one offset by $\simeq0\farcs1$ toward the outer  part of 
the disk with respect to the total intensity peak, located in the southwest region of the disk. This is reflected in Fig.
\ref{fig:poli}, especially for bands 6 and 7, where points with Stokes I brighter than $\sim 20$  and $\sim 
60$ \mJybeam~have larger polarized intensity levels than the mean polarization fraction. This is likely due to  the 
mismatch between the peak of polarized intensity and total intensity.

\section{Spectral properties of the dust emission}
\label{sec:si}

\begin{figure}[t!]
\includegraphics[width=\columnwidth]{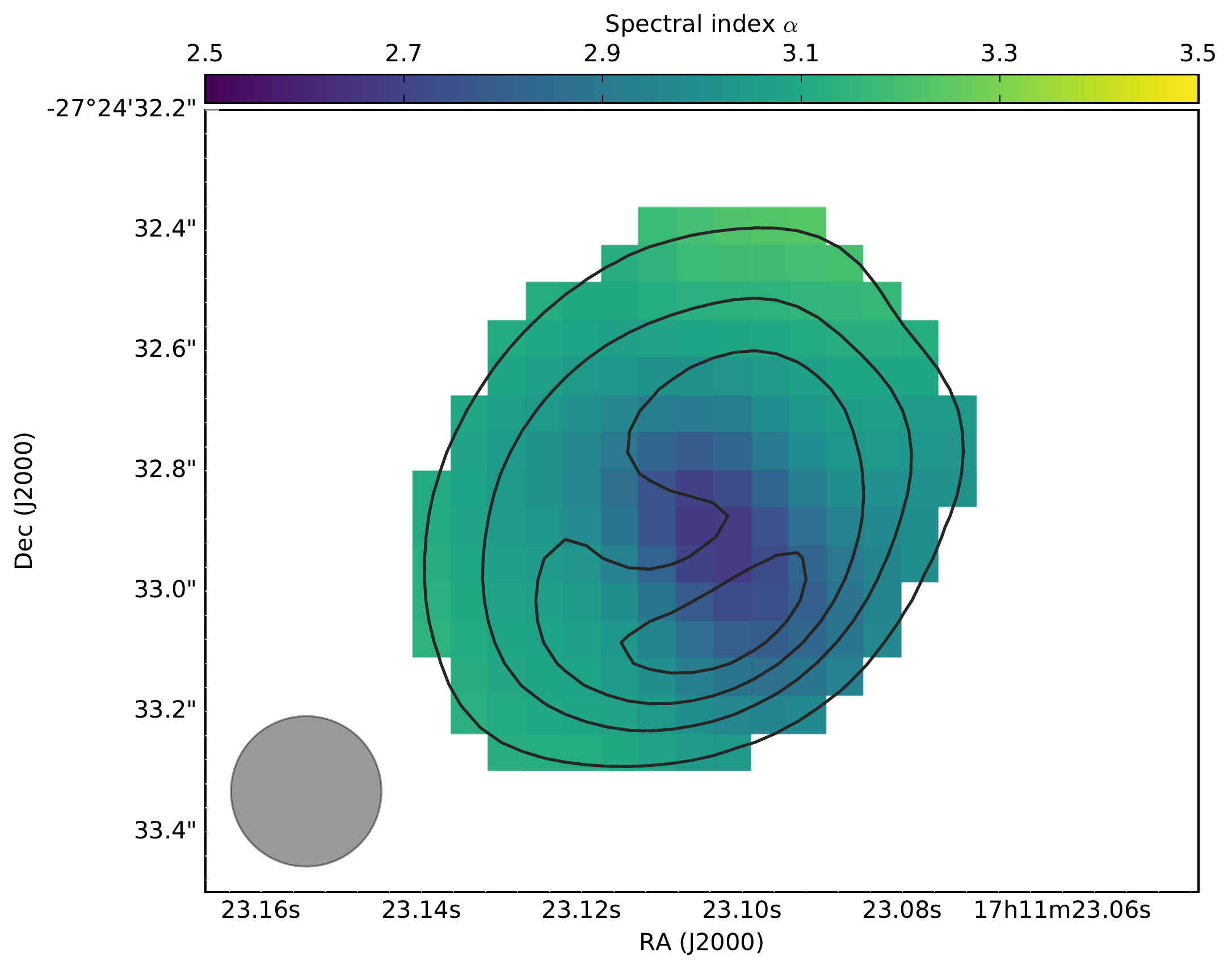}
\caption{
Spectral index map obtained from the Stokes I emission between bands 3  and 7. The Stokes I maps were smoothed to a 
common circular beam of 0.25$\arcsec$ prior to spectral index computation. The contours show the Stokes I Band 7 flux as in 
Fig. \ref{fig:cuts}. 
\label{fig:sindex}}
\end{figure}

Since Bands 3 and 7 have the largest frequency separation among the three bands, we have smoothed the 
corresponding Stokes I maps to a common beam and used their flux difference to compute the spectral index $\alpha$ as 
$S \propto \nu^{\alpha}$, where $S$ is the intensity and $\nu$ is the frequency.  Fig. \ref{fig:sindex} shows the spatial 
distribution of $\alpha$ with respect to Stokes I and polarized intensity. The spectral index has a mean value of $3.01 
\pm 0.02$ with a minimum $2.67 \pm 0.01$ near the center of the disk. We computed the dust opacity index $\beta$ 
assuming an optically thin and Rayleigh-Jeans regime for the disk edge ($\beta = \alpha - 2$), and found 
$\langle\beta\rangle \sim 1.0$, which is within typical values found in  in protostellar and YSO disks 
\citep[e.g.,][]{Natta07}. However, this value is lower than the value expected for the interstellar medium 
($\beta_{\mathrm{ISM}} \sim 1.7$) and suggests some grain growth, with maximum sizes in the millimeter range \citep{Natta07}. 

An accurate estimation of maximum grain size is affected by departures from the Rayleigh-Jeans regime and changes in 
optical depth through the disk. Although large, millimeter-size grains have been reported in other circumstellar disks 
\citep[e. g.,][]{Perez15}, we cannot discount the effect of optical depth in some regions of BHB07-11. In addition, young, 
Class I disks such as BHB07-11 are expected to have a population of grains significantly smaller than 1 mm, given that 
recent scattering models have inferred maximum grain sizes of $\sim 100$ $\mu$m in young (Class I/II) protoplanetary disks 
such as HL Tau \citep[e.g.][]{Kataoka16a}.

\section{Origin of polarization}
\label{sec:origin}

\subsection{Self-scattering of dust thermal emission?}
\label{subsec:self}

In the case of a population of grains with different sizes, as expected in the ISM and in protostellar disks, self-scattering is 
produced by the largest grains, and it has a strong wavelength dependence \citep{Kataoka15, Yang16a}. Thus, the 
maximum polarization degree is produced in a relatively narrow  band around a wavelength of $\lambda \sim  
2\pi a_{\mathrm{max}}$, where $a_{\mathrm{max}}$ is the maximum dust grain size \citep{Kataoka16b, Kataoka17}, 
since the polarization degree decreases significantly departing from this frequency for a specific $a_{\mathrm{max}}$. Thus for example 
a population of grains with maximum grain size of 150~$\mu$m would produce a averaged polarization degree above 
0.5\% in the dust emission between roughly 350~$\mu$m and 4~mm \citep[values obtained from Fig. 4 
of ][]{Kataoka17}.  
As observations and modeling have revealed so far, when self-scattering is the dominant mechanism the regions with strongest polarized intensity show typically small polarization levels,  <3\% \citep{Kataoka15,Yang17,Girart18}, although in the edges of the dust emission the polarization can increase to higher values \citep[e. g.,\,][]{Kataoka16b,Girart18}.
In addition, the self-scattering polarization pattern depends basically on the density distribution and disk geometry. The characteristic pattern of the polarization direction is mainly aligned with the disk minor axis 
\citep[e.g.][]{Kataoka16b, Stephens17,Hull18}. Optical depth effects can affect this pattern, but the mean direction is still aligned 
with the minor axis  \citep[e.g.][]{Yang17}. For an intensity distribution with lopsided pattern (such as a transition disk or 
ring-like disk), the polarization pattern has a clear radial component in the inner part of the ring and an azimuthal 
component in the outer part of the ring  \citep{Kataoka15, Kataoka16a}. This morphology was confirmed by ALMA 
observations of HD 142527, a Herbig Ae star surrounded by a protoplanetary disk \citep{Kataoka16b}. 
 
The observed polarization in the three ALMA bands does not fit the aforementioned properties for self-
scattering. First, the observed position angle pattern of the polarization is very different from the one expected for a 
lopsided disk \citep{Kataoka15, Kataoka16a}. Second, the polarization degree levels observed in our data (right panels of 
Fig. \ref{fig:cuts}) are at least a factor of $\sim 4$ (in Band 7, higher in Band 3) larger than the total polarization fraction 
predicted by self-scattering plus radiative alignment in a protoplanetary disk with $a_{\mathrm{max}} = 150$ $\mu$m, 
which is the grain size that contributes more to the modelled polarization \citep[see Fig. 4 in][]{Kataoka17}. Third, 
between band 3 and 7, the frequency has changed by a factor of 3.5, but the polarization degree has changed by a factor 
of 2.0, implying that the polarization degree changes more smoothly than expected in the self-scattering model. We 
therefore rule out self-scattering as potential mechanism to explain our polarization data. Indeed, if the maximum grain 
size derived from the spectral index analysis is correct ($a_{\mathrm{max}} \sim$ millimeter, see sect.~\ref{sec:si}), then the highest polarization from 
self-scattering should occur at wavelengths larger than 1~cm. But at 1~cm the dust peak emission for the measured spectral 
index would be 73~$\mu$Jy~beam$^{-1}$. Assuming a 3\% polarization degree for self-scattering the polarized intensity 
would be only few  $\mu$Jy~beam$^{-1}$, which would be only possible to detect with the Next Generation Very Large 
Array (ngVLA).
 
Although self-scattering is ruled out to explain our data, other potential mechanisms can still produce the observed polarization. We now discuss our results in the framework of grain alignment by radiative torques.
 
\subsection{Dust alignment with radiation anisotropy?}
\label{subsec:rat}

The dust polarization produced by radiation fields relies on radiative torques imparted on elongated dust grains by an 
anisotropic radiation field \citep{Lazarian07b}. This mechanism is efficient in aligning $\mu$m-size grains in the 
interstellar medium with the local magnetic field due to Larmor precession around magnetic field lines. The Larmor 
precession timescale is inversely proportional to the magnetic field strength and proportional to the dust effective area. 
This means that for large millimeter-size grains in circumstellar disks, the Larmor precession timescale  is longer than the 
gaseous damping timescale. As a result, the radiative precession of dust grains around the radiation direction will become more efficient and the dust grains will be 
aligned with the radiative flux rather than the magnetic field. 

In this scenario, since the alignment is between the radiation direction and the precession axis (which is perpendicular to 
the grain long axis), the polarization is perpendicular to the radiation flux gradient, and its overall distribution has a centrosymmetric morphology centered on the source of radiation \citep{Tazaki17}. In a full disk, the polarized pattern has an azimuthal 
orientation, while for a lopsided disk the polarization pattern is more complicated. However, if the radio sources revealed 
by the VLA observations are associated to a proto binary system, they could be a source of anisotropic radiation 
responsible for aligning grains in the disk. 

\subsection{Dust alignment with magnetic fields?}
\label{subsec:bfield}

If Larmor precession prevails over radiative precession, the polarization is produced by thermal emission of elongated grains aligned with the magnetic field \citep{Andersson15}.
It is well established that the polarization of the millimeter dust emission in dense molecular envelopes surrounding protostars and their disks trace the 
magnetic field, since the observed field pattern in some cases matches well the theoretical predictions \citep[see, 
e.\,g., ][]{Girart06, Girart09,Frau11}. Therefore, we speculate that being a Class I YSO, the dust grain population in the 
BHB07-11 disk is not too different from the surrounding envelope feeding the disk.

\begin{figure}[t!]
\includegraphics[width=\columnwidth]{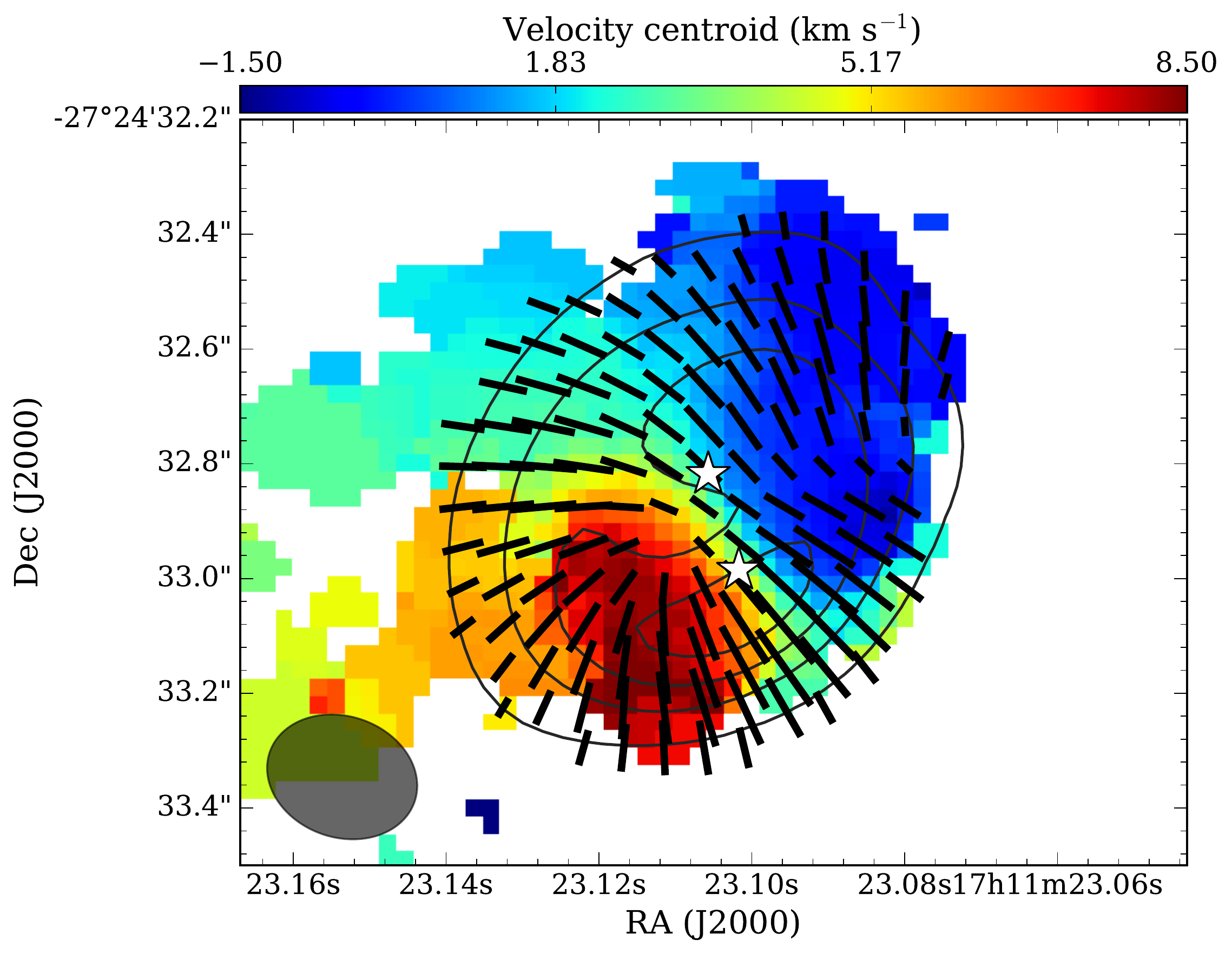}
\caption{Polarization vectors rotated by 90$\degr$ showing the plane-of-sky magnetic field configuration at Band 7.  
Contours show the same Stokes I intensity levels as in Fig. \ref{fig:cuts}. The background image is a moment 1 map of 
the H$_2$CO ($3_{2,1} - 2_{2,0}$) emission in BHB07-11. The moment map is produced from a 3$\sigma$ cut ($\sigma 
\sim 5.3$ mJy beam$^{-1}$) in the molecular emission. The synthesized beam of the H$_2$CO map is  $0.27\arcsec \times 0.21\arcsec$. The stars symbols have the same meaning as in Fig. \ref{fig:cuts}.
\label{fig:mag}}
\end{figure}

Upon the assumption that the polarization arises from grain alignment with the magnetic field, the 90\degr-rotated 
polarization map shows the plane-of-sky component of the disk magnetic (B) field averaged along the line of sight (Fig. 
\ref{fig:mag}). The vector map does not exhibit an obvious azimuthal morphology as previously reported in other young 
disks seen face-on \citep[e.\,g.,][]{Rao14}. Instead, the observed pattern could be interpreted as the dragging of the magnetic 
field lines due to the disk rotation, a precursor of a toroidal B-field morphology. As reported by \citet{Alves17}, the 
kinematics derived from molecular line observations of BHB07-11 is consistent with infall plus rotation motions at the 
inner envelope, but it is dominated by Keplerian rotation at scales of $\sim 150$ AU, corresponding essentially to the 
disk. Specifically, the H$_2$CO ($3_{2,1} - 2_{2,0}$) emission ($\nu$ = 218.760 GHz, E$_{u} \sim 70$ K) is confined to 
the disk in either low and high velocity components (which reach $\sim 5.5$ km s$^{-1}$ with respect to the systemic 
velocity of the object). The velocity centroid map of the molecular emission shows a velocity gradient along the disk long 
axis (Fig.\ref{fig:mag}). There is a clear correspondence between the putative B-field lines and the velocity structure. 

In the next section, we compare our data with polarization models of radiative alignment and 
magnetic alignment mechanisms.

\section{Polarization models}

\subsection{Alignment with radiation field}

\begin{figure}[t!]
\includegraphics[width=\columnwidth]{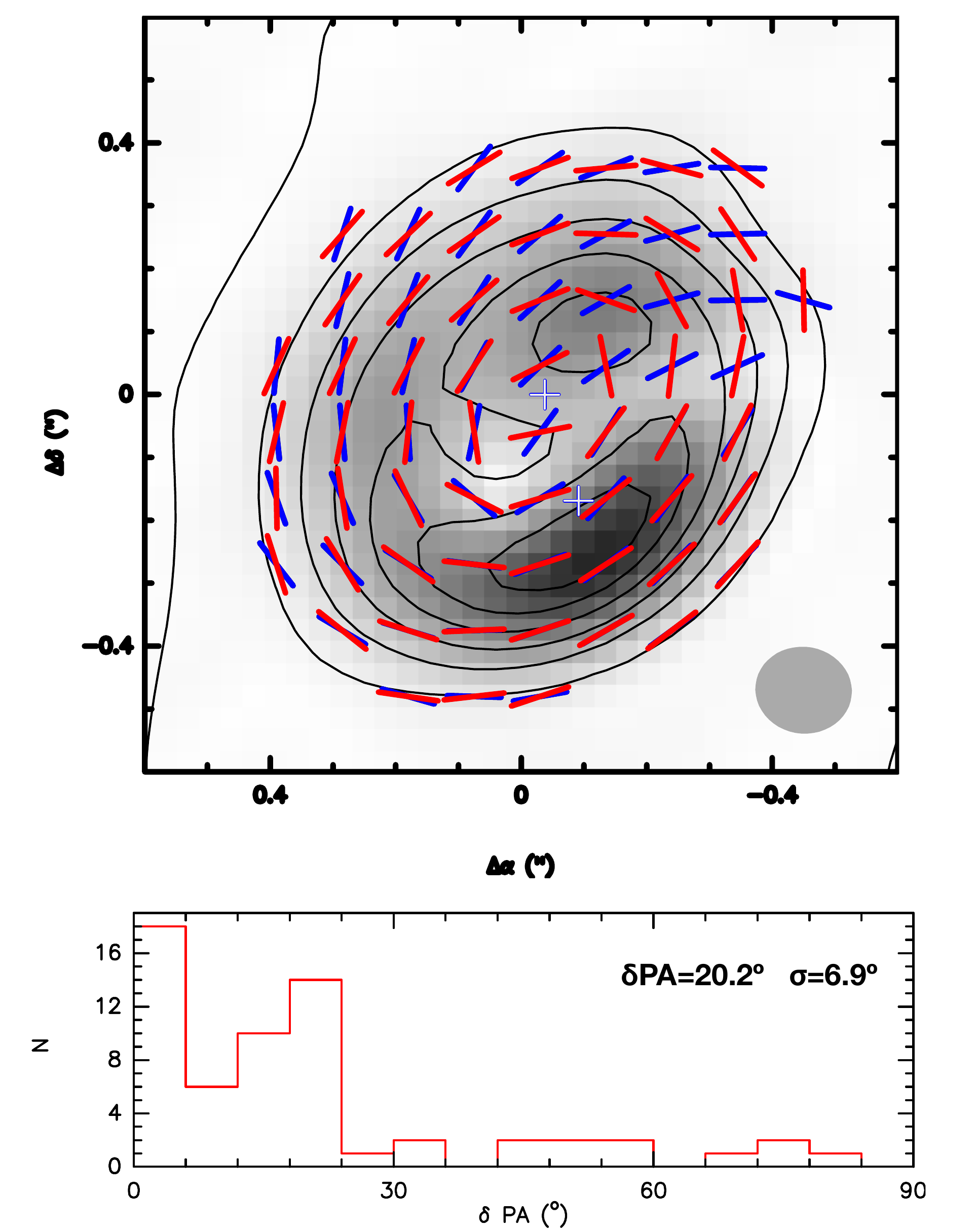}
\caption{Best fit for the polarization produced by radiation fields from the VLA 5a source (red vectors) and the Band 7 
data (blue vectors). The grey scale indicates polarized intensity with same levels of Fig. 1. Vectors are shown for every 3 pixels. The position angle difference between the our data and the 
radiation alignment position angles are shown as a histogram in the bottom panel. The histogram bin is 6\degr~to be 
consistent with the largest observational uncertainty in position angle.
\label{fig:rf}}
\end{figure}

We have built synthetic polarization maps assuming radiation alignment produced by the VLA 5a and VLA 5b sources. We have considered either polarization from each protostar individually and in combination weighted by $1/d^2$, where $d$ is the distance to the protostars. We used the disk inclination and position angle in the plane-of-sky as free parameters. For the scenario where VLA 5b is the main dominant source of radiation, the mean $\delta$PA= PA$_{\mathrm{data}}-$PA$_{\mathrm{model}}$ is 28.9$\degr$ with a standard deviation $\sigma$ of 5.5$\degr$. The mean $\delta$PA residuals are even larger (> 30$\degr$) for the model assuming combined radiation from the two protostars. The best fit was achieved in the scenario where VLA 5a is the dominant source of radiation, from which a disk inclination of 48\degr~and position angle of 138\degr~were derived (fit error of $\pm 15\degr$). These values are consistent with our observational estimates \citep{Alves17}. Fig. \ref{fig:rf} shows the comparison between our Band 7 polarization maps and the radiation fields polarization produced by VLA 5a, with mean $\delta$PA$= 20.2\degr$ and $\sigma = 6.9\degr$. However, there is a strong discrepancy in position angles at 
the northwest portion of the disk. The histogram in Fig. \ref{fig:rf} shows the distribution of residuals ($\delta$PA) with bins 
the size the largest observational uncertainty in position angle ($\sigma_{\mathrm{PA}}^{\mathrm{max}} = 6\degr$). 

\begin{figure*}[t!]
\includegraphics[width=\textwidth]{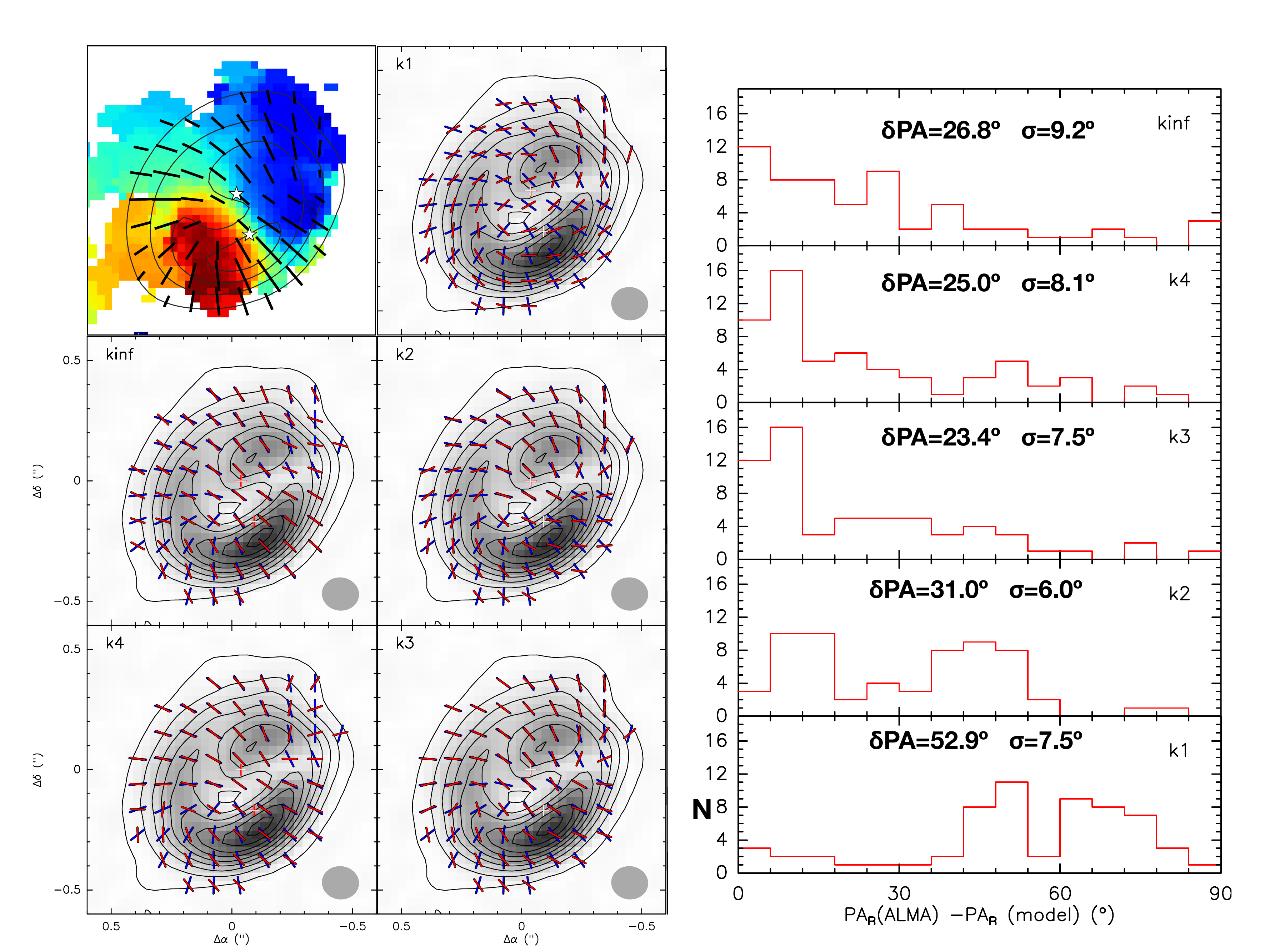}
\caption{Left panels show synthetic magnetic field lines (red vectors) plotted on top of observed Band 7 Stokes I contours (5 to 45 mJy beam$^{-1}$ in steps of 5 mJy beam$^{-1}$), polarized intensity (grey scale with same intensity levels of Fig. 1) 
and magnetic fields (blue vectors, which are 90$\degr$-rotated with respect to Fig. 1). The red vectors result from our
rotating disk modeling assuming different $k = B_p/B_\phi$ values. Fig. \ref{fig:mag} is shown in the upper left panel 
for reference. The right panels show the residuals from each case, with mean $\delta$PA = PA$_{\mathrm{data}}-$PA$_{\mathrm{model}}$ values and standard deviation $\sigma$ indicated on each histogram panel.
\label{fig:bmodel}}
\end{figure*}

\subsection{Alignment with magnetic field}

For the case of magnetic alignment, we used the DustPol module of the ARTIST software 
\citep{Padovani12, Jorgensen14} to build a disk model including thermal continuum polarization by dust grains aligned by 
a large-scale magnetic field. In order to model the configuration of the observed magnetic field lines, we assumed the 
model by \citet{Shu07} of a viscously  accreting disk surrounding a low-mass protostar, with magnetic fields that are 
radially advected by accretion and azimuthally twisted by differential rotation. At steady state, the inward advection of field 
lines is balanced by the outward diffusion associated to the turbulent viscosity of the disk, and the large-scale field 
reaches a configuration characterized by an inclination with respect to the disk normal of $i=50-60\degr$ and a poloidal-
to-toroidal ratio $k=B_p/B_\phi$ of order unity. In our modelling, we adopted $i=55$\degr and considered $k$ as a free 
parameter.

We ran DustPol assuming a frequency of 343 GHz as in Band 7 observations. The disk model has the normal to the disk
plane inclined by 
52\degr\,with respect to the line of sight and it is then rotated by 138\degr\,in the plane of the sky (East of North). DustPol creates a set of FITS files of the Stokes parameters that can be straightforwardly used as an 
input for the {\tt simobserve/simanalyze} tasks of CASA. For these tasks we assumed an antenna configuration 
corresponding to the observations carried out in Band 7.

In Fig. \ref{fig:bmodel} we show the fit results for the 90$\degr$-rotated polarisation produced by thermal 
dust emission for $k$ values ranging from 1 ($k1$, poloidal and toroidal magnetic field components are equally strong) to
infinite $k$ ($kinf$, purely poloidal field). The best fit is represented by the $k=3$ case, with mean
residual $\delta$PA$= 23.4\degr$ and standard deviation $\sigma=7.5\degr$. This case corresponds to a toroidal component a third of the poloidal one. The extreme cases of purely poloidal and strong toroidal field do not fit the data. Our results are consistent 
with the rotation seen in our molecular line maps, where the sense of rotation is the same as the sense of the B-field 
twist. The existence of the poloidal component indicates that the field is being advected inward by an accretion flow. 
Indeed, this is a reasonable interpretation because, as a young source, infall gas motions and outflow ejection are 
expected to happen and, indeed, were reported through extended H$_2$CO ($3_{0,3} - 2_{0,2}$) emission and CO 
($2-1$) lines in \citet{Alves17}.

\subsection{Interpreting the modelling results}

As shown by Fig. \ref{fig:rf} and \ref{fig:bmodel},  the differences between the observed polarization angles and the model 
predictions for radiative alignment and magnetic alignment are similar. In both cases the agreement is generally good, but 
some discrepancies remain between our data and the synthetic maps for some regions of the disk. Although formally the 
radiative alignment model gives a slightly better fit of the polarization angles than the magnetic alignment model, interpreting 
our results in terms of the former model presents some difficulties: the observed polarization levels are significantly higher 
than predicted by \citet{Tazaki17}, and the pattern is more spiral-like than centrosymmetric. In the context of the combined
models of self-scattering plus radiative alignment reported in \citet{Kataoka17}, in order to have significant polarization
by radiative alignment, one would need grains with sizes of 150~$\mu$m. This would imply significant self-scattering polarization 
in Band 7. However, since we do not see any signature of self-scattering pattern in our maps, and the observed polarization
levels are much higher than the model predictions, it seems unlikely that the polarization arises from radiative alignment. 
More generally, we do not see the apparent wavelength dependence of the polarization pattern in the different bands, 
reported in \citet{Kataoka17} and \citet{Stephens17}.

On the other hand, as long as Larmor precession dominates over radiative precession, the polarization pattern  by
grains aligned with the magnetic field is not expected to vary with wavelength, as shown by our maps. In addition, the twisting 
of the magnetic field required to fit our data is in the same sense as the rotation revealed by the molecular line centroid map.  
The discrepancy with the data could be explained in this case by the peculiar nature of the disk:  the VLA data indicate a 
binary system embedded in the disk, which possibly implies the presence of gaps and tiny circumstellar disks
around each stellar component. Such an environment is expected to produce substructure in the magnetic field morphology that 
could distort the large-scale magnetic field components.

Although we cannot rule out alignment with the radiation field, our analysis shows that the observed polarization is 
consistent with being produced by magnetically aligned grains. The middle and right panels of Fig. \ref{fig:cuts} show that the 
polarized radiation is emitted from zones near the outer parts of the disk, where small grains are more abundant. This is an 
indication that polarization usually traces the external disk layers, which explains the similar polarization patterns in all three 
bands. In this scenario, the larger millimeter-size grains suggested by our spectral index analysis of the Stokes I emission are 
located in the disk mid-plane.

\section{Conclusions}

We have performed multi-frequency ALMA polarization observations toward the young circumbinary disk in BHB07-11. 
Our observations revealed extended and ordered polarization in all three bands (3, 6 and 7). Our main conclusions 
are:
\begin{itemize}
\item While the total and polarized intensity increases with frequency, the polarization morphology is the same across the 
three bands. This gives strong support against polarization mechanisms such as self-scattering and radiation fields, 
whose predictions are wavelength dependent;
\item In spite of the asymmetry observed in BHB07-11 and its inclination with respect to the line-of-sight, the polarization structure does not match the self-scattering predictions for lopsided and inclined disks. In addition, the observed polarization levels are at least a factor of 4 larger than model predictions for scattering polarization in protoplanetary disks 
(however, similar levels of polarization observed in other objects have been attributed to self-scattering);
\item Our data do not match models of radiative alignment assuming the flux of VLA 5b or the weighted combined flux of VLA 5a and VLA 5b as sources of radiation, but they match relatively well the predictions if VLA 5a alone is assumed the main source of radiation. However, we cannot explain the discrepancy in the context of this mechanism since the observed spiral-like polarization morphology is inconsistent with the centrosymmetric predictions, and the observed polarization levels are  higher than the model predictions.
\item Our maps are consistent with a model of a rotating disk with a poloidal magnetic field component plus a toroidal 
component produced by the disk rotation. The synthetic magnetic field lines fit the sense of rotation derived from molecular line maps of the disk. We thus conclude that the polarization in our data is produced by magnetically aligned dust grains.
\end{itemize}

\begin{acknowledgements}
 We would like to thank the anonymous referee for a constructive report. This paper makes use of the following ALMA data: ADS/JAO.ALMA\#2013.1.00291.S and 
 ADS/JAO.ALMA\#2016.1.01186.S. ALMA is a partnership of ESO (representing its member states), NSF (USA) and 
 NINS (Japan), together with NRC (Canada), MOST and ASIAA (Taiwan), and KASI (Republic of Korea), in 
 cooperation with the Republic of Chile. The Joint ALMA Observatory is operated by ESO, AUI/NRAO and NAOJ. JMG is  
 supported by the MINECO (Spain) AYA2014-57369-C3 grant. MP acknowledges funding from the European Unions
  Horizon 2020 research and innovation programme under the Marie Sk\l{}odowska-Curie grant agreement No 664931. 
  G.A.P.F. acknowledges the partial support from CNPq and FAPEMIG (Brazil). PC acknowledges support of the 
  European Research Council (ERC, project PALs 320620). WV acknowledges support from the ERC through 
  consolidator grant 614264. The data reported in this paper are archived in the ALMA Science Archive. 
  \end{acknowledgements}

%
%

\bibliographystyle{aa}
\bibliography{refalves}

\end{document}